\documentclass[prd,showpacs,preprintnumbers,twocolumn,amsmath,nofootinbib,amssymb]{revtex4}
\usepackage{graphicx,color,dcolumn,booktabs,bm}
\usepackage{graphics}
\usepackage{longtable,lscape}
\usepackage{txfonts}
\usepackage{overpic}
\usepackage{amssymb}
\usepackage{epstopdf}
\usepackage{indentfirst}
\usepackage{feynmf}   
\usepackage{slashed}  
\usepackage{cases}
\usepackage{color}
\usepackage{float}
\usepackage{multirow}
\usepackage{graphicx,color,dcolumn,booktabs,bm}
\usepackage{epsfig,dsfont,amssymb,amsmath,amsfonts,amsbsy,mathrsfs}

\graphicspath{{Figures/}} %
\usepackage[colorlinks, citecolor=blue,anchorcolor=red,menucolor=red, linkcolor=red,filecolor=red,runcolor=red,urlcolor=blue,frenchlinks=red]{hyperref}
\makeatletter
\@addtoreset{equation}{section}
\makeatother

\allowdisplaybreaks
\begin{document}
\title{${Q} \bar{Q}$ $({Q}\in{\{b,c}\})$ spectroscopy using the modified Rovibrational model}
\author{Zheng-Yuan Fang$^{1,2}$}\email{fang1628671420@163.com}
\author{Ya-Rong Wang$^{1,2}$}\email{nanoshine@foxmail.com}
\author{Cheng-Qun Pang$^{1,2,3}$\footnote{Corresponding author}}\email{pcq@qhnu.edu.cn}
\affiliation{$^1$College of Physics and Electronic Information Engineering, Qinghai Normal University, Xining 810000, China\\$^2$Joint Research Center for Physics,
Lanzhou University and Qinghai Normal University,
Xining 810000, China \\$^3$Lanzhou Center for Theoretical Physics, Key Laboratory of Theoretical Physics of Gansu Province, Lanzhou University, Lanzhou, Gansu 730000, China}
\begin{abstract}
Mass spectra of quarkonium systems can be described by different phenomenological potentials. In the present work, the {\color{black}resonance} states of heavy quarkonium like ($c\bar{c}$ and $b\bar{b}$) are considered as the rovibrational states. We study a parameterized rovibrational model derived from the empirical solution of the nonrelativistic Schr\"{o}dinger equation with Morse potential, the corrections are composed of colour hyperfine interaction and spin-orbit interaction of mesons. We {\color{black}obtain} the high excited state mass spectra of charmonium and bottomonium comparing the results in reasonable agreement with the present experimental data.

\end{abstract}
\maketitle

\section{Introduction}\label{sec1}
{\color{black}{Since the discovery of charm quarks and bottom quarks in 1974 and  1977, respectively}}, heavy quarkonium has become an influential and attractive research field because its physical processes cover the whole energy range of Quantum Chromodynamics (QCD). This energy range provides us an excellent place to study the properties of perturbative and non-perturbative QCD \cite{aubert1974experimental,herb1977observation,innes1977observation}. With the improvement of experimental accuracy, {\color{black}Particle Data Group (PDG)} has collected dozens of heavy quarkonium particles. The study of hadronic energy spectrum is one of the main tasks at present. We classify it and then study higher radial and orbital excited states of {\color{black}heavy quarkonium}. In previous work, a series of phenomenological potential models can use specific techniques to describe the mass spectra of quarkonium and are in good agreement with the experimental values \cite{fischer2015spectra,PhysRevD.74.014012,PhysRevD.75.074031,soni2018q,ebert2011spectroscopy,PhysRevD.95.034026,PhysRevD.79.094004,PhysRevD.86.034015,Bai_Qing_2009}. Abundant experimental data can accurately fit different parametric potential models, and these models are consistent in the scale range of observable states by solving the Schr\"{o}dinger equation. For example, Barnes et al, used the standard coulomb potential and linear potential to calculate the energy spectrum and electromagnetic radiation decay of charmonium \cite{barnes2005higher}. Considering the color screening effect, Jun-Zhang Wang et al, calculated the energy spectrum and two-body strong decays of $c\bar{c}$ and $b\bar{b}$ states, respectively \cite{wang2019constructing,wang2018higher}. The meson family high excited state spectroscopy describes the most familiar Reggie trajectory \cite{ebert2013spectroscopy,chen2018regge,chen2021structure}. In particular, the latest discussion on the overall framework of hadronic spectrum also provides a basis for our calculation \cite{albaladejo2021novel}.
\par
Due to the abundant experimental data of heavy mesons, the rovibrational model derived from the empirical solution of the nonrelativistic Schr\"{o}dinger equation with morse potential is only used to describe the excited spectrum of hadrons rather than to calculate the effective potential between quarks. Bernardo and Bastos calculated the mass of partial $\pi$, $K$, $N$, $\Sigma$ orbital and radial resonance states, which were in good agreement with the experimental results \cite{bernardo2021hadron}. Rovibrational model takes into account the effect of colour hyperfine interaction and spin-orbit interaction, and has more systematic description of quark model. In the model, the hadronic family has the resonance states with the same isospin type corresponding to the fitting parameters of the same set of experimental particles, and we further predict the mass spectra of other high excited hadrons.
\par
In 1966, Paulie noticed that more than half of the baryon resonance can be described as the vibration and rotation of the semi-rigid rotor \cite{pauling1966baryon}. Bernardo and Bastos consider the non-harmonic and semi-rigid rotation correction of the vibrational rotation. However, when there are many families of hadron resonances, the predicted values are somewhat different from those predicted by the traditional quark model. In this work, we consider the rovibrational model with spin-orbit coupling, spin-spin coupling and tensor items to calculate the mass spectra of charmonium and bottomonium, where $Y(4220)$ is calculated as $\psi(4^3S_1)$ \cite{wang2019constructing,PhysRevD.95.034026,PhysRevD.86.034015,PhysRevD.79.094004,barnes2005higher}. Our work is better in comparison with available experimental than other theoretical values.
\par
This paper is aimed to give a {\color{black}{systemic study of excited states of charmonium and bottomonium}} mass spectroscopy. Verify the feasibility of the model. Morse potential method of diatomic molecular mass spectrum is mature, after correction to calculate the double heavy quark particle mass spectrum also achieved coincident with the experimental.
\par
This paper is organized as follows. In Sec. \ref{sec2}, the models employed in this work are briefly reviewed.
The mass spectrum of charmonium and bottomonium will be performed in {\color{black}{Sec. \ref{sec3}}. The paper ends with a conclusion in Sec. \ref{sec4}.

\vspace{-5mm}
\section{Models employed in the work}\label{sec2}

\vspace{-1mm}
In this work, the modified rovibrational model is utilized to calculate the mass spectrum of the charmonium and bottomonium meson family. A comparison of calculated rovibrational states with the experimental and other phenomenological potential models results is predicted. In the following, these models will be illustrated briefly.

\vspace{-1mm}
\subsection{The modified rovibrational model}

In 1929, the exact solutions of the Schr\"{o}dinger equation for the motion of atomic nuclei in diatomic molecules are given \cite{morse1929diatomic}. It is noteworthy that approximate solutions of Schr\"{o}dinger equation with some diatomic molecular interactions using Nikiforov-Uvarov(NU) method can obtained the energy eigenvalues and the total normalized wave function \cite{okon2017approximate}. Hesham Mansour et al. choose a phenomenological potential in the framework of the
nonrelativistic Schr\"{o}dinger equation with relativistic corrections using the Nikiforov-Uvarov (NU) method obtained the meson mass spectroscopy \cite{mansour2022meson}. The energy level behavior of diatomic molecules can be described by the stationary Schr\"{o}dinger equation with the internuclear potential in the form of morse potential, which considers the rotation-vibration coupling including anharmonic and semi-rigid rotor corrections \cite{bernardo2021hadron}. {\color{black} In case of charmonium and bottomonium meson, the quark and antiquark separated by a distance $r$
}.
The radial equation is given as follows:
\begin{equation}
\begin{split}
\frac{d^2R(r)}{dr^2}+\frac{L(L+1)R(r)}{r^2}=\frac{-8\pi^2m}{h^2}[E-D_{e}-D_{e}e^{-2\alpha(r-r_{e})}\\
+2D_{e}e^{-\alpha(r-r_{e})}]R(r),\label{2.1}
\end{split}
\end{equation}
where $R(r)$ is the radial wave function, $E$ is the non-relativistic total energy of the system, $h$ is the Planck's constant, $m$ is the mass of the particle, $L$ is the orbital quantum number, $D_{e}-D_{e}e^{-2\alpha(r-r_{e})}+2D_{e}e^{-\alpha(r-r_{e})}$ is the popular morse potential, $D_{e}$ and $\alpha$ are parameters characteristic to each particle. $r_{e}$ is equated with the equilibrium internuclear separation for diatomic molecules. The morse potential will be used
here to describe only the excited spectrum of hadrons and not to calculate the effective potential of the quarks. Comparison of Cornell potential model and Godfrey–Isgur (GI) model for calculating charmonium and bottomonium mass spectrum with considered colour hyperfine interaction and spin-orbit in
teraction. Using the Bernardo and Bastos
solution \cite{bernardo2021hadron}, which includes the rovibrational coupling and corrections composed of spin-orbit coupling term, spin-spin coupling term and tensor term for states with $L>0$, the mass of the resonances $M(n,L)$ can be calculated as
\begin{equation}
\begin{split}
M(n,L)=D_{e}+\hbar\omega(n+\frac{1}{2})-\hbar\omega\chi_{e}(n+\frac{1}{2})^2+B_{rot}L(L+1)\\-D_{rot}L^2(L+1)^2-\alpha_{e}(n+\frac{1}{2})L(L+1)
+\frac{1}{2}a_{e}(S(S+1)-\frac{3}{2})\\+\frac{1}{2}b_{e}(J(J+1)-S(S+1)-L(L+1))+c_{e}\boldsymbol{T},\label{2.2}
\end{split}
\end{equation}
where $n$ is the radial quantum number, $\hbar\omega(n+\frac{1}{2})$ is the harmonic contribution, $\hbar\omega\chi_{e}(n+\frac{1}{2})^2$ is the harmonic correction, $B_{rot}L(L+1)$ is the rigid rotor contribution, $D_{rot}L^2(L+1)^2$ represents the semi-rigid rotor correction, and the rovibrational coupling is given by $\alpha_{e}(n+\frac{1}{2})L(L+1)$, which is the Coriolis effect. Consideration of orbital and radial high excited states in the charmonium and bottomonium system. $\frac{1}{2}a_{e}(S(S+1)-\frac{3}{2})$ and $c_{e}\boldsymbol{T}$ represents the colour  hyperfine interactions correction, which the GI nonrelativistic quark model, there are typical forms of contact interaction and tensor interaction \cite{godfrey1985mesons}. $\frac{1}{2}b_{e}(J(J+1)-S(S+1)-L(L+1))$ denotes the spin-orbit interaction due to one-gluon exchange correction.
$\boldsymbol{S_{Q}}$ and $\boldsymbol{S_{\bar{Q}}}$ are the spins of quarks in meson component, and the total spin is $\boldsymbol{S}=\boldsymbol{S_{Q}}+\boldsymbol{S_{\bar{Q}}}$. The spin-orbit operator is diagonal in a $|\boldsymbol{J,S,L}\rangle$ basis, spin-dependent terms with
the matrix elements includes
\begin{align}
\langle\boldsymbol{S_{Q}}\cdot\boldsymbol{S_{\bar{Q}}\rangle}=\frac{1}{2}(S(S+1)-\frac{3}{2}).
\label{2.3}
\end{align}
 We can get the relationship between spin and orbital coupling by $6j$ symbol, the spin-orbit
interaction correction terms with the matrix elements includes
\begin{align}
 \langle\boldsymbol{L\cdot{S}}\rangle=\frac{1}{2}(J(J+1)-S(S+1)-L(L+1)).
\end{align}
   The tensor operator $\boldsymbol{T}$ has nonvanishing
diagonal matrix elements only between $L > 0$ spin-triplet
states, $\boldsymbol{r}$ is any unit vector. which are \cite{barnes2005higher}\\
 \begin{align}
 \boldsymbol{T}=(\frac{\boldsymbol{S_{Q}\cdot{r}S_{\bar{Q}}\cdot{r}}}{r^2}-\frac{1}{3}\boldsymbol{S_{Q} \cdot S_{\bar{Q}}}),\\
 \langle^3L_{J}|\boldsymbol{T}|^3L_{J}\rangle=\left\{\begin{array}{l}-\frac{1}{6(2 L+3)}, J=L+1; \\ +\frac{1}{6}, J=L; \\ -\frac{L+1}{6(2 L-1)}, J=L-1.\end{array}\right.
\end{align}
The parameter $B_{rot}=\frac{\hbar^2}{2mr_{e}^2}$ determined using the experimental mass correlation particles, $D_{e}$ and coefficients $\hbar\omega$, $\hbar\omega\chi_{e}$, $D_{rot}$, $B_{rot}$, $\alpha_{e}$, $a_{e}$, $b_{e}$, $c_{e}$ as a function are determined by fitting the corresponding hadrons system experimental results of the radial and orbital quantum numbers, respectively. Due to the calculation of highly excited hadron spectrum, and the lowest mass ground state has significant influence on the system. It is noteworthy that the ground state $\eta_{c}(1^1S_{0})$ of charmonium and the ground state $\eta_{b}(1S)$ of bottomonium do not participate in the fitting of the high excited state results more in line with the experimental values. In this case, in order to make the whole fitting systematic, we choose to fit both with the orbital and radial excited states that have been observed experimentally. The behavior of the rovibrational parameters of Eq. (\ref{2.2}) is similar to that of diatomic molecule states, and for the meson system with double heavy quarks, the theoretical results obtained after considered the influence of the colour hyperfine interaction and spin-orbit interaction of mesons on the high excited state was better agreement with the experimental results. This is also the original intention to plus these perturbations. For molecules and hadrons, the behavior parameters of the corresponding meson system depend on the experimental results of the system. For $L=0$, there is no rotation effect, the contribution of rovibrational model are mainly concentrated in $D_{e}$ and the harmonic contribution, the anharmonic correction and the perturbation effects are relatively weak. For $L>0$, which overall calculation results of the hadrons system are in agreement with the experimental data.
\section{Numerical results and phenomenological analysis} \label{sec3}
 We have determined the free parameters of the model by minimizing the quantity
 \begin{eqnarray}
 \chi^2=\sum\limits_{i}(\frac{E^{th}_{i}-M^{exp}_{i}}{M^{exp}_{i}})^2.
 \end{eqnarray}
The parameters of charmonium fitted from fifteen experimental values. Experimentally observed mass values of bottomonium
which are used to calculate the minimu $\chi^2$ and to obtain the model parameters listed in the third column of Table \ref{SGIfit1}. Applying the modify rovibrational model and the parameters in Table \ref{SGIfit1}, the mass spectra of the charmonium and bottomonium can be obtained, as shown in Table \ref{massc} and Table \ref{massb}.

\vspace{-0.5cm}
\renewcommand{\arraystretch}{1.2}
\begin{table}[htbp]
\caption{Parameters of the modified rovibrational model, which are determined by fitting the charmonium and bottomonium system experimental data listed in PDG  averages. \label{SGIfit1}}
\begin{center}
\begin{tabular}{cccc}
\toprule[0.8pt]\toprule[0.8pt]
Parameter & Value $(c\bar{c})$  &Value $(b\bar{b})$   \\
\midrule[0.8pt]
$D_e$ (MeV) &2712.52&9171.35\\
$\hbar\omega$ (MeV) &767.05&597.70\\
$\hbar\omega\chi_{e} $ (MeV) &95.43&48.89 \\
$B_{rot}$ (MeV) &273.02&306.82\\
$D_{rot}$ (MeV)&21.20&25.54\\
$\alpha_e\ $(MeV)&39.12&69.77\\
$a_e$ (MeV)&40.90 &8.55\\
$b_e$ (MeV)&12.27 &13.07\\
$c_e$\ (MeV)&134.42 &26.81\\
\bottomrule[0.8pt]\bottomrule[0.8pt]
\end{tabular}
\end{center}
\end{table}

\vspace{-2cm}
\subsubsection{The spectra of charmonium meson excitations}
 The spectra of charmonium meson orbital and radial excitations are calculated, and the values are listed in {Table} \ref{massb}. All experimental masses are taken from PDG averages \cite{Zyla:2020zbs}.
 The third radial excited state of $\eta_c(3S)$  has a mass of 4.003 GeV, which is {\color{black}{smaller}} than the result of GI model and close to that reported in Ref \cite{barnes2005higher}. The difference between theoretical results for radial excited states $\eta_c(3S)$ and $\eta_c(4S)$ are about 0.2 GeV. For the ground  state of $\psi(1^3D_1)$ close to the experimental values, its second and third radial excited states ($\psi(2^3D_1)$ and $\psi(3^3D_1)$) have the mass {of 4.122 GeV and  4.273 GeV}, respectively. When evaluating the theoretical results of all states, the qualitative evaluation errors selection of all states are $\mathcal{V}^{Er}_{i}$=5.0 MeV, which are larger than their respective experimental uncertainty. The reason is that the experimental error of these particles is relatively small and the distribution is unevenly. Finally, we obtain that error of the whole system is $\chi^2/{N}$  is 55.8, which is consistent with the similar theory model \cite{PhysRevD.86.034015}. Our work is better in
comparison with available experimental than other values in the literatures \cite{fischer2015spectra,PhysRevD.74.014012,soni2018q,ebert2011spectroscopy,PhysRevD.95.034026,PhysRevD.79.094004,PhysRevD.86.034015,barnes2005higher,wang2019constructing,
mansour2022meson,shah2012leptonic,Patel_2016}.

\begin{table*}[htp]
\caption{The mass spectra of the experimental missing orbital and radial excited states have been predicted. Here number of experimental values ($N=15$), we list the error analysis is $\chi^2=$$\sum\limits_{i}$$(\frac{E^{th}_{i}-M^{exp}_{i}}{\mathcal{V}^{Er}_{i}})^2$, where $Y(4220)$ is calculated as $\psi(4^3S_1)$ state. The values of underline in $\psi(4^3S_1)$ state denote the predicted results of different phenomenological models after fitting the adopt experimental data of underline in parentheses, respectively. The parentheses in the reciprocal second column show the relative error between the theoretical and experimental values after fitting each experimental data of charmonium in the model.
All results are in units of MeV. \label{massb}}
\vspace{-20pt}
\setlength{\tabcolsep}{1.5mm}{
\begin{center}
\[\begin{array}{cccccccccccccccc}
\toprule[0.8pt]\toprule[0.8pt]
 \text{State}  & \text{This work}&\text{NU \cite{mansour2022meson}}&\text{\cite{ebert2011spectroscopy}}&\text{\cite{PhysRevD.86.034015}}&\text{\cite{PhysRevD.95.034026}} &\text{\cite{soni2018q}} &\text{\cite{PhysRevD.79.094004}}&\text{\cite{PhysRevD.75.074031}}&\text{\cite{PhysRevD.86.034015}} &\text{\cite{PhysRevD.74.014012}}&\text{\cite{Patel_2016}}&\text{JZ \cite{wang2019constructing}} &\text{GI    \cite{barnes2005higher}}&\text{PDG \cite{Zyla:2020zbs}}&\text{Error}\\
 \midrule[0.8pt]
 \eta_c(1^1S_0)      & {3041}& 3041 & 2930&2981&2984&2989&2979&2980&2980&3088&2979& 2981& 2982& 2984\ (1.9$\%$)    & 5.0  \\
 \psi(1^3S_1)      & {3075}& 3140 & 3110&3096&3097&3094&3097&3097&3097&3168&3096& 3096& 3098& 3097\ (0.71$\%$)  & 5.0  \\
 \eta_c(2^1S_0)    & {3618}& 3661 & 3680&3635&3637&3602&3623&3597&3633&3669&3600& 3642& 3623& 3638\ (0.55$\%$)   & 5.0  \\
 \psi(2^3S_1)      & {3651}& 3702 & 3680&3685&3679&3681&3673&3685&3690&3707&3680& 3683& 3676& 3686\ (0.95$\%$) & 5.0  \\
 \eta_c(3^1S_0)    & {4003}& 4135 &  -  &3989&4004&4058&3991&4014&3992&4067&4011& 4013& 4064& -              &      \\
 \psi(3^3S_1)      & {4037}& 4050 & 3800&4039&4030&4129&4022&4095&4030&4094&4077& 4035& 4100& 4039\ (0.05$\%$)    & 5.0  \\
 \eta_c(4^1S_0)    & {4197}& 4414 & -  &4401&4264&4448&4250&4433&4244&4398&4397& 4260& 4225&  -              &      \\
 \psi(4^3S_1)      & {4231}& $\underline{4419}$ & -  &$\underline{4427}$&4281&$\underline{4514}$&4273&$\underline{4477}$&4273&$\underline{4420}$&$\underline{4454}$& 4274& 4225& 4230\  ($\underline{4421}$)       & 5.0  \\
  h_c(1^1P_1)      & {3467}& 3518 & 3430&3525&3526&3470&3519&3527&3524&3536&3536& 3538& 3517& 3525\ (1.65$\%$)  & 5.0  \\
 \chi_{c0}(1^3P_0) & {3435}& 3414 & 3320&3413&3415&3428&3433&3416&3392&3448&3488& 3464& 3445& 3415\ (0.58$\%$)   & 5.0  \\
 \chi_{c1}(1^3P_1) & {3512}& 3504 & 3490&3511&3521&3468&3510&3508&3491&3520&3514& 3530& 3510& 3511\ (0.03$\%$)  & 5.0  \\
 \chi_{c2}(1^3P_2) & {3515}& 3489 & 3550&3555&3553&3480&3556&3558&3570&3564&3565& 3571& 3550& 3556\ (1.15$\%$)  & 5.0  \\
  h_c(2^1P_1)      & {3962}& 3824 & 3750&3926&3916&3943&3908&3960&3922&3950&3996& 3933& 3956&  -              &      \\
 \chi_{c0}(2^3P_0) & {3933}& 3765 & 3830&3870&3848&3897&3842&3844&3845&3870&3947& 3896& 3916& 3918\ (0.38$\%$)   & 5.0  \\
 \chi_{c1}(2^3P_1) & {4013}& 3808 & 3670&3906&3914&3938&3901&3940&3902&3934&3972& 3929& 3953& -              &      \\
 \chi_{c2}(2^3P_2) & {4010}& 3915 &  -  &3949&3937&3955&3937&3994&3949&3976&4021& 3952& 3979& 3927\ (2.11$\%$)   & 5.0  \\
  h_c(3^1P_1)      & {4269}& 4137 &  -  &4337&4139&4344&4184& -  &4137&4291& -  & 4200& 4318&  -              &      \\
 \chi_{c0}(3^3P_0) & {4240}& 4080 & -  &4301&4146&4296&4131& - &4192&4214& -  & 4177& 4292&  -              &      \\
 \chi_{c1}(3^3P_1) & {4320}& 4121 & 3910&4319&4192&4338&4178& -  &4178&4275& -  & 4197& 4317&  -              &      \\
 \chi_{c2}(3^3P_2) & {4317}& 4151 & -  &4354&4211&4358&4208   &4212&4316& -  & 4213& 4292& -              &      \\
  h_c(4^1P_1)      & {4385}& 4416 &  -  &4744& -  &4704& -  & -  & -  & -  & -  & 4389&  -  & -              &      \\
 \chi_{c0}(4^3P_0) & {4356}& 4362 &  -  &4698& -  &4653& -  & -  & -  & -  & -  & 4374&  -  &  -              &      \\
 \chi_{c1}(4^3P_1) & {4436}& 4401 & -  &4728& -  &4696& -  & - & -  & -  & -  & 4387& -  & -              &      \\
 \chi_{c2}(4^3P_2) & {4434}& 4230 & -  &4763& -  &4718& -  & -  & -  & -  & -  & 4398&  -  & -              &      \\
 \eta_{c2}(1^1D_2) & {3799}& 3478 & 3807&3807&3805&3765&3796&3824&3802&3803&3796& 3848& 3837&  -              &      \\
 \psi({1^3D_1})    & {3781}& 3402 & 3739&3783&3792&3775&3787&3804&3729&3789&3792& 3830& 3819& 3778\ (0.08$\%$)    & 5.0  \\
 \psi_2({1^3D_2})  & {3850}& 3461 & 3550&3795&3807&3772&3798&3824&3788&3804&3794& 3848& 3838& 3822\ (0.73$\%$)    & 5.0  \\
 \psi_3({1^3D_3})  & {3861}& 3514 & 3869&3813&3808&3755&3799&3831&3844&3809&3798& 3859& 3849&  -              &      \\
 \eta_{c2}(2^1D_2) & {4141}& 3828 &  -  &4196&4108&4182&4099&4191&4105&4158&4224& 4137& 4208& -              &      \\
 \psi({2^3D_1})    & {4122}& 3756 &  -  &4105&4095&4188&4089&4164&4057&4143&4222& 4125& 4194& 4159\ (0.88$\%$) & 5.0  \\
 \psi_2({2^3D_2})  & {4192}& 3812 & -  &4190&4109&4188&4100&4189&4095&4159&4223& 4137& 4208& -              &      \\
 \psi_3({2^3D_3})  & {4203}& 3863 & -  &4220&4112&4176&4103&4202&4132&4167&4425& 4144& 4217&  -              &      \\
 \eta_{c2}(3^1D_2) & {4291}& 4141 &  -  &3549&4336&4553&4326& -  &4330& -  & -  & 4343&  -  & -              &      \\
 \psi({3^3D_1})    & {4273}& 4072 & - &4507&4324&4555&4317& -  &4293& -  & -  & 4334&  -  &  -             &      \\
 \psi_2({3^3D_2})  & {4342}& 4125 & -  &4544&4337&4557&4327& -  &4322& -  & -  & 4343&  -  &  -              &      \\
 \psi_3({3^3D_3})  & {4354}& 4174 &  -  &4574&4340&4549&4331& -  &4351& -  & -  & 4348& -  & -              &      \\
 \midrule[1pt]
\chi^2/{N}         & 55.8     & 1250   &  & 17.5 &  33.1  &  89.4  &  38.5  & 49.7   &  61.1  & 72.4   & 64.4     & 28.8&29.1 &\\
\bottomrule[0.8pt]\bottomrule[0.8pt]
\end{array}\]
\end{center}}
\end{table*}

\begin{table*}[htp]
\caption{The mass spectra of bottomonium mesons, here number of experimental values ($N$=18), we list the error analysis is $\chi^2=$$\sum\limits_{i}$$(\frac{E^{th}_{i}-M^{exp}_{i}}{\mathcal{V}^{Er}_{i}})^2$. The parentheses in the reciprocal second column show the relative error between the theoretical and experimental values after fitting each experimental data of bottomonium in the model. All results are in units of MeV.}
 \label{massc}
\vspace{10pt}

\setlength{\tabcolsep}{3mm}{
\begin{tabular}{ccccccccccc}
    \toprule[0.8pt]\toprule[0.8pt]
\text{State}&$n^{2S+1}L_J$&\text{This work}&\text{NU \cite{mansour2022meson}}&\text{\cite{PhysRevD.92.054034}}&\text{\cite{PhysRevD.93.074027}}&\text{\cite{Bai_Qing_2009}}&\text{JZ \cite{wang2018higher}}&\text{GI \cite{barnes2005higher}}&\text{PDG \cite{Zyla:2020zbs}}&\text{Error}\\
 \midrule[1pt]
$\eta_b(1S)$& $1^1S_0$     &{9451}  &9436 &9402  &9455 &9389& 9398 & 9394 &9399 (0.55$\%$)&5.0   \\
 $\eta_b(2S)$& $2^1S_0$     &{9951}  &9991 &9976  &9990 &9987& 9989 & 9975 &9999 (0.48$\%$)&5.0   \\
 $\eta_b(3S)$& $3^1S_0$     &{10354} &10139&10336 &10330&10330& 10336& 10333&--&5.0   \\
 $\eta_b(4S)$& $4^1S_0$     &{10658} &10324&10523 &--   &10595& 10597& 10616&--&5.0   \\
 $\eta_b(5S)$& $5^1S_0$     &{10865} &10498&10869 &--   &10817& 10597& 10860&--&5.0   \\
 $\eta_b(6S)$& $6^1S_0$     &{10973} &10662&11097 &--   &11011& 10991& 11079&--&5.0   \\
 $\Upsilon(1S)$& $1^3S_1$   &{9458}  &9491 &9465  &9502 &9460& 9463 & 9459 &9460 (0.02$\%$) &5.0 \\
 $\Upsilon(2S)$& $2^3S_1$   &{9958}  &10013&10003 &10015&10016& 10017& 10004& 10023 (0.65$\%$)&5.0 \\
 $\Upsilon(3S)$& $3^3S_1$   &{10361} &10328&10354 &10349&10351& 10356& 10354& 10355 (0.06$\%$)&5.0 \\
 $\Upsilon(4S)$& $4^3S_1$   &{10665} &10546&10635 &10607&10611& 10612& 10633& 10579 (0.81$\%$)&5.0 \\
 $\Upsilon(10860)$&$5^3S_1$ &{10872} &10826&10878 &10818&10831& 10822& 10875& 10882 (0.09$\%$)&5.0 \\
 $\Upsilon(11020)$&$6^3S_1$ &{10980} &10971&11102 &10995&10988& 11001& 11092& 11003 (0.2$\%$)&5.0 \\
 $\Upsilon(7S)$  & $7^3S_1$ &{10991} &--   &--    &--   &--   & 11157& 11294&--&5.0 \\
 $h_b(1P)$& $1^1P_1$        &{9893}  &9879 &9882  &9879 &9903& 9894 & 9881 &9899 (0.06$\%$)&5.0 \\
 $h_b(2P)$&$ 2^1P_1$        &{10254} &10222&10250 &10240&10256& 10259& 10250&10260 (0.06$\%$)&5.0 \\
 $h_b(3P)$& $3^1P_1$        &{10516} &10417&10541 &10516&10529& 10530& 10540&--& \\
 $h_b(4P)$& $4^1P_1$        &{10681} &10324&10790 &--   &10757& 10751& 10790&--& \\
 $h_b(5P)$& $5^1P_1$        &{10748} &10498&11016 &--   &10955& 10938& 11013&--& \\
 $\chi_{b0}(1P)$&$1^3P_0$   &{9867}  &9843 &9847  &9855 &9865& 9858 & 9845 &9859 (0.08$\%$)&5.0 \\
 $\chi_{b0}(2P)$&$2^3P_0$   &{10227} &10196&10226 &10221&10226& 10235& 10225&10231 (0.04$\%$)&5.0 \\
 $\chi_{b0}(3P)$&$3^3P_0$   &{10490} &10134&10552 &10500&10502& 10513& 10521&--& \\
 $\chi_{b0}(4P)$&$4^3P_0$   &{10655} &10319&10775 &--   &10732& 10736& 10773&--& \\
 $\chi_{b0}(5P)$&$5^3P_0$   &{10722} &10494&11004 &--   &10933& 10926& 10998&--&\\
 $\chi_{b1}(1P)$&$1^3P_1$   &{9893}  &9874 &9876  &9874 &9897& 9889 & 9875&9893 (0$\%$)&5.0 \\
 $\chi_{b1}(2P)$&$2^3P_1$   &{10254} &10217&10246 &10236&10251& 10255& 10246&10256 (0.01$\%$)&5.0 \\
 $\chi_{b1}(3P)$&$3^3P_1$   &{10516} &10138&10538 &10513&10524& 10527& 10537&10512 (0.04$\%$)&5.0 \\
 $\chi_{b2}(1P)$&$1^3P_2$   &{9914}  &9891 &9897  &9886 &9918& 9910 & 9896&9912 (0.01$\%$)&5.0 \\
 $\chi_{b2}(2P)$&$2^3P_2$   &{10274} &10230&10261 &10246&10269& 10269& 10261&10269 (0.05$\%$)&5.0 \\
 $\chi_{b2}(3P)$&$3^3P_2$   &{10537} &10141&10550 &10521&10540& 10539& 10549&--\\
 $\chi_{b2}(4P)$&$4^3P_2$   &{10702} &10325&10798 &--   &10767& 10758& 10797&--\\
 $\chi_{b2}(5P)$&$5^3P_2$   &{10767} &--   &--    &--   &--   & 10944& 11020&--\\
 $\Upsilon(1D) $&$1^3D_2$   &{10164} &10113&10147 &10122&10151& 10162& 10147&10164 (0$\%$)&5.0 \\
 $\Upsilon_2(2D) $&$2^3D_2$ &{10245} &9943&10449  &10418&10438& 10450& 10449&--&\\
 $\Upsilon_1(1D) $&$1^3D_1$ &{10198} &9729&10138  &10117&10145& 10153& 10137&--&\\
 $\Upsilon_1(2D) $&$2^3D_1$ &{10210} &9938&10441  &10414&10432& 10442& 10441&--&\\
 $\Upsilon_3(1D) $&$1^3D_3$ &{10198} &9739&10115  &10127&10156& 10170& 10155&--&\\
 $\Upsilon_3(2D) $&$2^3D_3$ &{10279} &9947&10455  &10422&10442& 10456& 10455&--&\\
 \midrule[0.8pt]
 $\chi^2/{N}$ &  & $38.8$ & 351.6 &35.7& $32.4$ &10&  11.3&31.4& &\\
 \bottomrule[0.8pt]\bottomrule[0.8pt]
\end{tabular}}
\end{table*}

\begin{figure}[H]
\centering
\includegraphics[height=5cm,width=7cm]{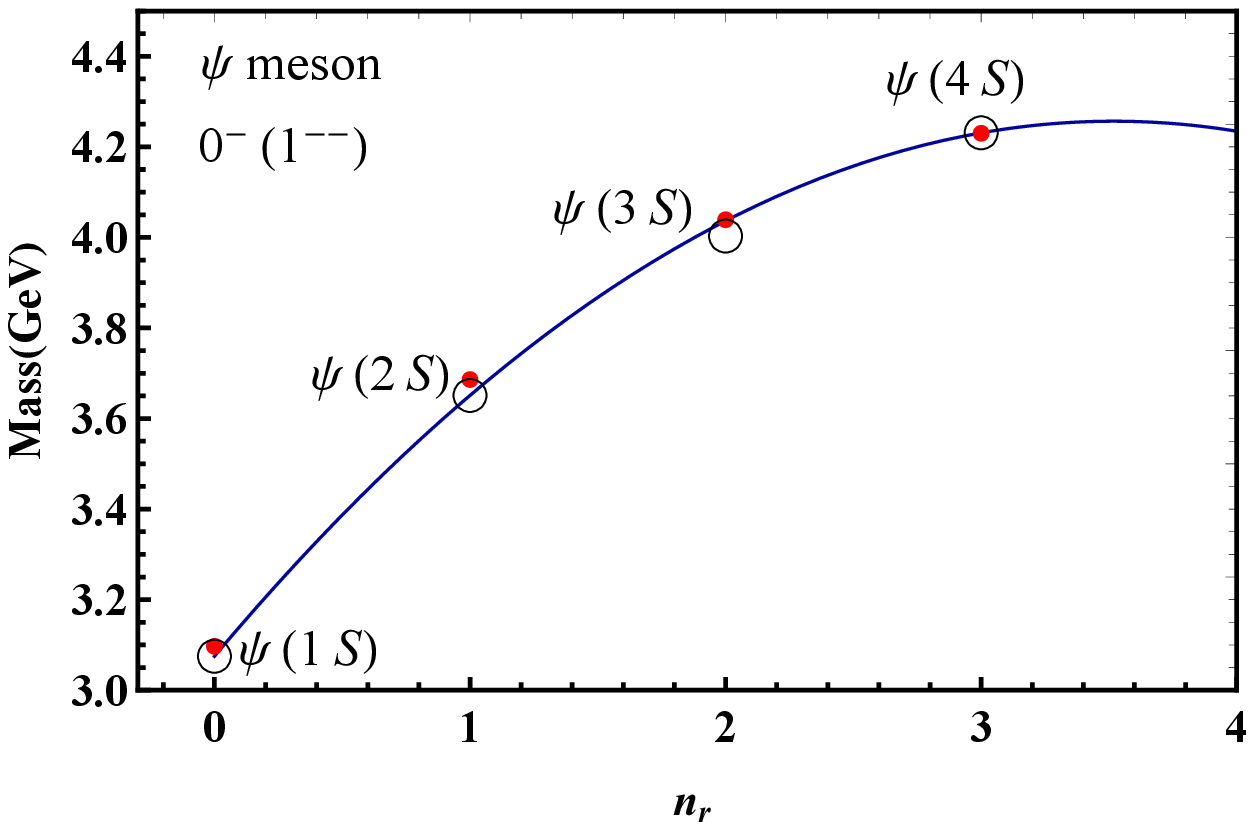}
\label{pic1}
\end{figure}

\vspace{-0.7cm}
\begin{figure}[H]
\centering
\includegraphics[height=5cm,width=7cm]{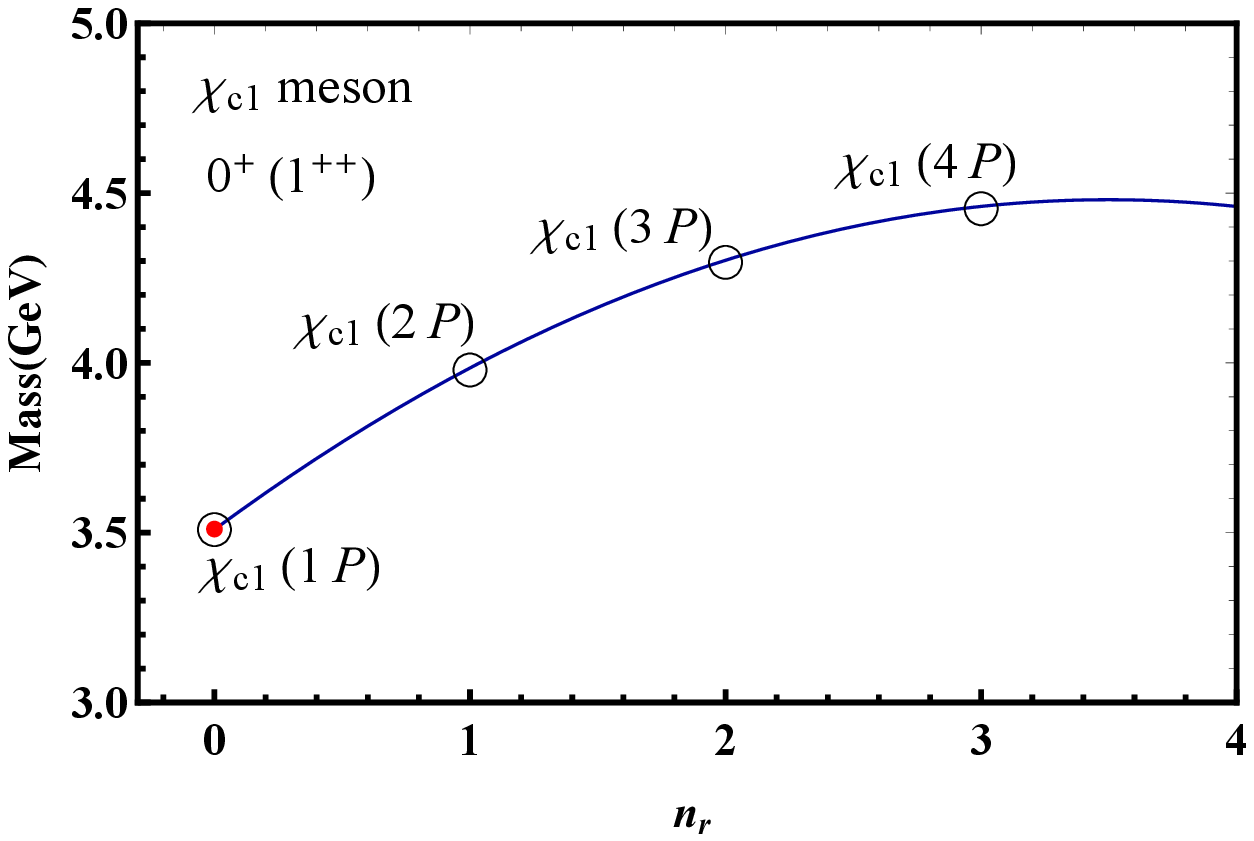}
\label{pic2}
\end{figure}

\vspace{-0.7cm}
\begin{figure}[H]
\centering
\includegraphics[height=5cm,width=7cm]{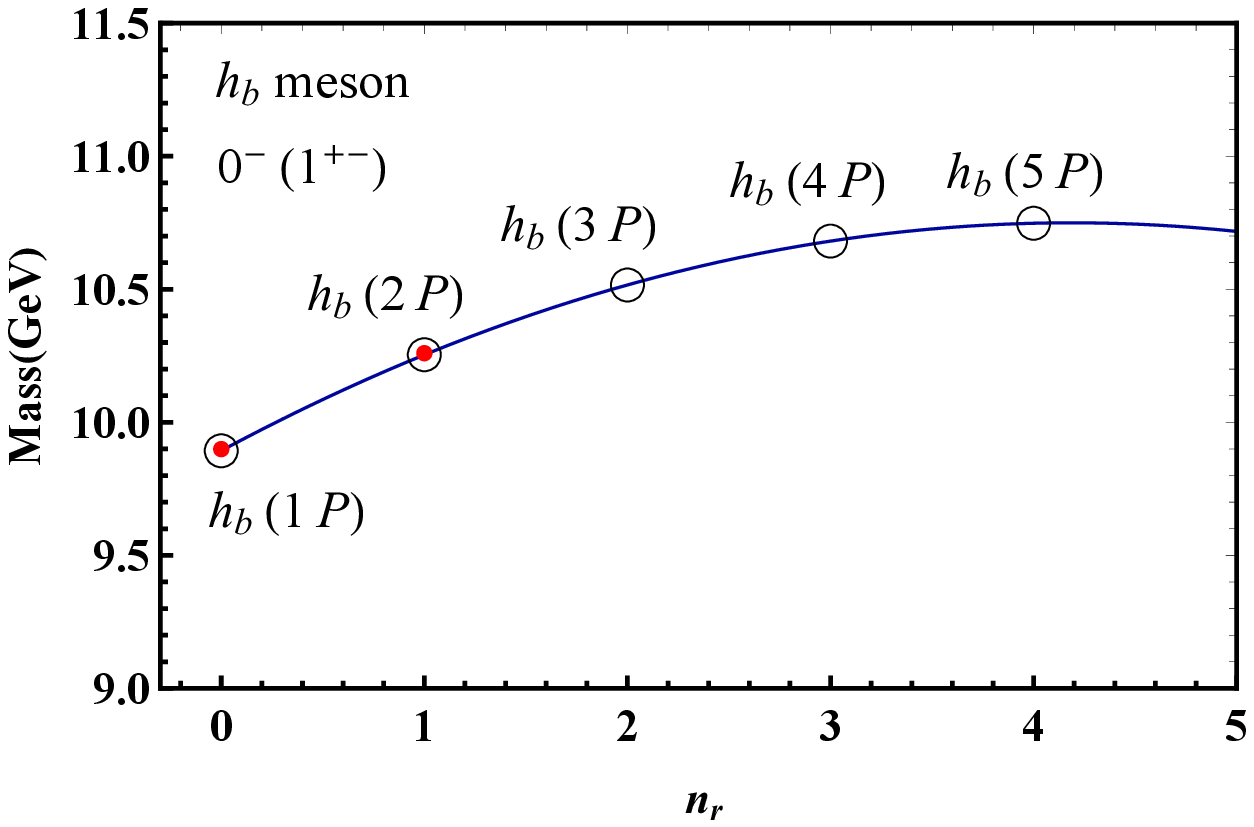}
\label{pic3}
\end{figure}

\vspace{-0.7cm}
\begin{figure}[H]
\centering
\includegraphics[height=5cm,width=7cm]{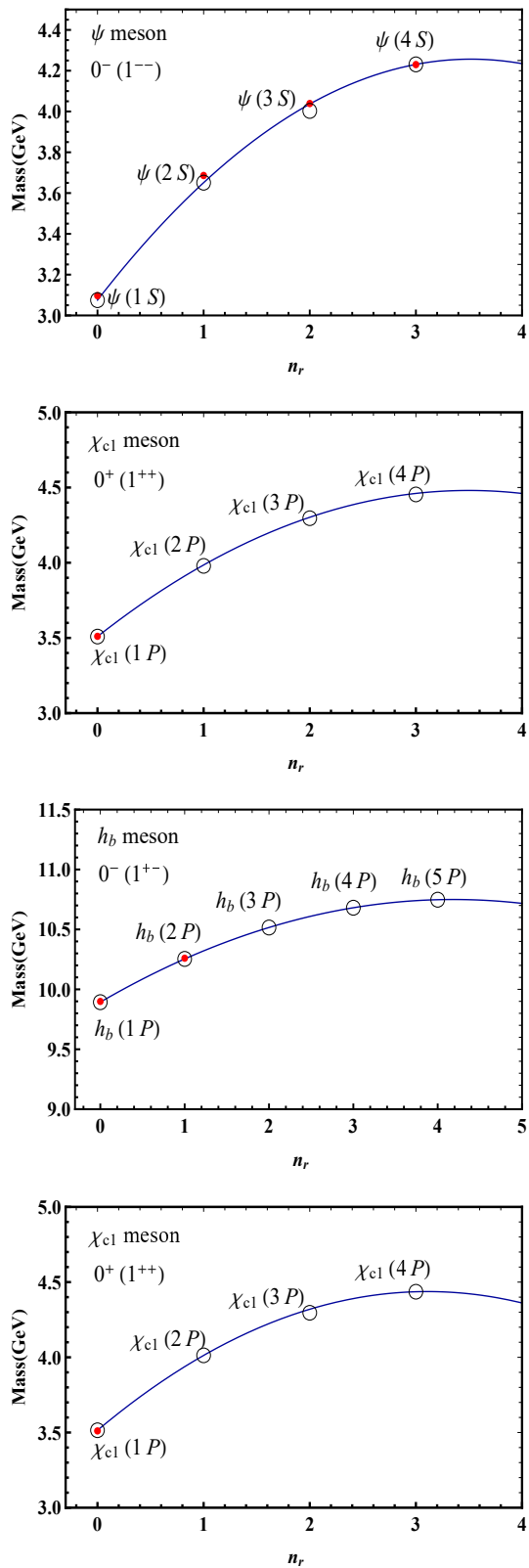}
\label{pic4}
\vspace{-0.3cm}
\caption{Fitted curves of the $\psi$, $\chi_{c1}$, $h_b$, $\chi_{b2}$ radial $(M,n)$ resonances calculated by the modified rovibrational model, here the radial quantum number is $n = n_r + 1$, solid-line curve show clearly the mass variation trend when different radial quantum number $n$ is taken. The open circle and the filled geometry are the theoretical and experimental values, respectively.}
\end{figure}

\subsubsection{The spectra of bottomonium meson excitations}
 {Table} \ref{massc} show that have been established theoretical and experimental values. The results of the thirteenth column are calculated using the modified GI relativistic quark model with screening effects in bottomonium system \cite{wang2018higher}, all experimental masses are taken from PDG averages \cite{Zyla:2020zbs}. Although GI model successfully studied the bottomonium spectrum, the qualitative evaluation errors selection of all states are $\mathcal{V}^{Er}_{i}$=5.0 MeV, comparing $\chi^2/{N}$ values, one can easily see that the fitted modified rovibrational model improves the whole description of the bottomonium spectrum. Since the screening potential can lower the energy of the high excited states. In the words, the modified GI model well improves the accuracy of the whole bottomonium system. The calculation results of other models are listed for comparison in the literatures \cite{Bai_Qing_2009,barnes2005higher,wang2018higher,mansour2022meson,PhysRevD.92.054034,PhysRevD.93.074027}.

%

\section{conclusion}\label{sec4}
This paper presents analysis of mass spectra of the radial and orbital resonances of charmonium and bottomonium system, using the modified rovibrational model. Rovibrational model study on diatomic molecular spectra show satisfactory agreement with experimental values for many molecules. Until now, the rovibrational model is useful for classifying the hadron resonances, especially for charmonium and bottomonium system with double heavy quarks. The modified Godfrey-Isgur quark model and Cornell potential model with the screening effect traditionally include spin-spin interaction and orbital-spin interaction. The rovibrational model of diatomic molecular mass spectra includes the vibration and rotation effects of two atoms, so it is spin-independent. For small scale of hadrons, spin and orbital dependence must be considered. According to the comparison of the fitted curves of the $\psi$, $\chi_{c1}$, $h_b$, $\chi_{b2}$ radial $(M,n)$ resonances calculated by the modified rovibrational, the calculation of too high excited state energy spectrum has some limitations. The reason is that the interaction of morse potential decreases at high excited states. It is noteworthy that many hadronic molecules are mainly ground states, and the modified rovibrational model has certain reference significance.

\section{ACKNOWLEDGMENTS}

This work is supported  by the National Natural Science Foundation of China under Grants No. 11965016, the projects funded by Science and Technology Department of Qinghai Province (No. 2020-ZJ-728).
\bibliographystyle{apsrev4-1}

\bibliography{hepref}

\end{document}